%% file: main.tex
\documentclass{llncs}
\usepackage{geometry}
\usepackage{listings,xcolor,url,pgfplots,pgfgantt,extarrows,tikz-3dplot,multirow,amsfonts}
\usepackage{pgfplots}
\usepackage[utf8]{inputenc} 
\usepackage[caption=false]{subfig}
\usepackage[T1]{fontenc}

\lstdefinelanguage{CAL}%
{classoffset=0,
	morekeywords={action,actor,else,end, function, priority, schedule, fsm, guard, var, do, if,not,then,network,structure},
	classoffset=1,
	morekeywords={true, false,{==>}},
	classoffset=0,%
	basicstyle=\fontsize{6}{6}\selectfont\sffamily,
	numberstyle=\tiny, 
	breaklines=true,
	backgroundcolor=\color{gray!10},
	tabsize=1,\textbf{}
	numbersep=0.5pt,
	xleftmargin=.01pt,
	xrightmargin=.0pt,
	morecomment=[s]{/*}{*/},%
	morecomment=[l]{//},%
	moredelim=[is][\color{DarkGoldenrod}]{|}{|},
	morestring=[b]",%
	morestring=[b]',
}[keywords,comments,strings]

\def\BibTeX{{\rm B\kern-.05em{\sc i\kern-.025em b}\kern-.08em
    T\kern-.1667em\lower.7ex\hbox{E}\kern-.125emX}}

\begin{document}
\title{Secure-by-design smart contract based on dataflow implementations}
%
%
\author{
Simone Casale-Brunet\inst{1,2}\protect\thanks{Dr. Simone Casale Brunet, PhD: \email{simone@casalebrunet.com}} \and
Marco Mattavelli\inst{3}\protect\thanks{Dr. MER Marco Mattavelli, PhD: \email{marco.mattavelli@epfl.ch}}}

%
\institute{ 
  Casale Brunet Consulting, Switzerland \and
  Blockchain Research Lab, Germany  \and
  \'{E}cole Polytechnique F\'{e}d\'{e}rale de Lausanne, Switzerland
}

\authorrunning{S. Casale-Brunet et al.}

%
\maketitle              

\input{tex/abstract}
\input{tex/introduction}
\input{tex/history}

\input{tex/atm}
\input{tex/dao_code}
\input{tex/bestpractices}
\input{tex/dataflow}

\input{tex/conclusions}
\footnotesize
\bibliographystyle{splncs04}

\end{document}

%% file: tex/abstract.tex
\begin{abstract}
This article conducts an extensive examination of the persisting challenges related to smart contract attacks within blockchain networks, with a particular focus on the reentrancy attack. It emphasizes the inherent vulnerabilities embedded in the programming languages commonly employed for smart contract development, particularly within Ethereum Virtual Machine (EVM)-based blockchains. While the concrete example used primarily employs the Solidity programming language, the insights garnered from this study are readily generalizable to a wide array of blockchain architectures.
Significantly, this article extends beyond the mere identification of vulnerabilities and ventures into the realm of proactive security measures. It explores the adaptation and adoption of dataflow programming paradigms, employing Domain-Specific Languages (DSLs) to enforce security by design in the context of smart contract development. This forward-looking approach aims to bolster the foundational principles of blockchain security, offering a promising research direction for mitigating the risks associated with smart contract vulnerabilities.
The objective of this article is to cater to a diverse audience, ranging from individuals with limited computer science and programming expertise to seasoned experts in the field. It provides a comprehensive and accessible resource for fostering a deeper understanding of the intricate dynamics between blockchain technology and the imperative need for secure smart contract development practices.
\end{abstract}

%% file: tex/introduction.tex
\section{Introduction}
\label{s:into}

Smart contracts (SCs) were initially conceptualized in the early 1990s as digital agreements characterized by automated enforcement and execution of legally binding terms. In recent years, they have been integrated into blockchain technology. However, critical bugs and vulnerabilities in SCs have led to catastrophic consequences for deployed applications, necessitating further scientific research to enhance their security and reliability.
Blockchain technology's foundational principle is the immutability and irreversibility of recorded data. This characteristic makes modifying deployed SCs infeasible, often requiring the creation of entirely new SCs for rectification. Rigorous pre-deployment testing and validation of SCs are crucial due to the impracticality of this approach. Unfortunately, contemporary testing methodologies often fall short, resulting in errors and vulnerabilities with severe repercussions.
A significant challenge arises from the disparity between the programming languages used in SC development and the unique characteristics of blockchain systems. Current programming techniques for smart contracts primarily rely on serial models, making it challenging to express parallel and distributed execution on a blockchain network. This limitation, described in recent papers such as~\cite{2020_Huashan} (see for example insights 4 and 13) and~\cite{2017_Ilya}, contributes to the difficulty of achieving secure and correct-by-construction smart contract implementations.

Presently, SC implementation often relies on vague and underspecified "coding best practices" to compensate for these shortcomings. This deficiency stems from the absence of a suitable design methodology and the unreliability of analysis and verification tools. The question of "which programming model best suits SCs?" remains an open scientific problem, as highlighted in previous research~\cite{2017_Ilya,2020_Sarwar,2020_Huashan,2022_Heidelinde}. Previous attempts to identify an effective programming model have yielded limited results, with existing literature primarily offering diverse solutions (often lacking scientific verification) tailored for specific and straightforward use cases~\cite{2022_Heidelinde}.

The aim of this work is to investigate the root causes of these vulnerabilities and challenges in achieving secure-by-design development techniques, which we attribute to the use of seemingly inappropriate programming languages. In fact, by examining a simple yet well-known example of a reentrancy attack (currently the most common and with the most catastrophic outcomes), it is possible to mitigate such vulnerabilities by adopting a dataflow programming model. This programming model, which has been successfully applied in fields such as video coding and genomics, enables the representation of a program (in this case, the smart contract) at a high level while ensuring both security and efficiency when implemented on parallel and heterogeneous architectures (in this case, the blockchain).

The paper is structured as follows:
Section~\ref{s:history} provides a brief historical overview of reentrancy attacks. It illustrates their emergence in the early days of Ethereum in 2016 and, despite the years that have passed, how reentrancy attacks remain a prevalent threat in newly deployed smart contracts, emphasizing the urgency for effective mitigation strategies. An illustrative scenario involving a bank ATM is presented in Section~\ref{s:atm} to facilitate understanding for non-technical readers.
Successively, Section~\ref{s:technical_analysis} delves deep into the technical aspects of reentrancy attacks, offering a detailed examination of the source code of a real-world smart contract. Through this analysis, readers can gain a profound understanding of why current programming languages for smart contract development are inadequate and fraught with risks. The limitations of current best practices are discussed in Section~\ref{s:bestpractices}.
Finally, the concept of a dataflow model for secure-by-design smart contract implementation is discussed in Section~\ref{s:dataflow}. Here, it is outlined how this model can be leveraged to construct inherently secure smart contracts. By doing so, the paper presents a potential solution to address the prevailing issues in the current ecosystem, where the development process heavily relies on the developer's experience rather than robust engineering methodologies.
Section~\ref{s:conclusions} concludes the paper, providing further research directions.


%% file: tex/history.tex
\section{The DAO Hack and how the history was altered with a fork}
\label{s:history}
In 2015, the nascent Ethereum community initiated discussions surrounding the concept of Decentralized Autonomous Organizations (DAOs). These blockchain-based entities were designed to facilitate coordinated human activities through the execution of verifiable code, primarily by utilizing smart contracts on the Ethereum blockchain. They aimed to enable decentralized decision-making regarding community protocols. In 2016, approximately one year after the Ethereum mainnet's launch, a DAO called "The DAO" was established. It operated as a decentralized, community-managed investment fund, with its smart contract deployed on April 30, 2016. Individuals acquired The DAO's community tokens by depositing Ether (ETH), and these ETH holdings constituted the investment funds managed by The DAO on behalf of its token-holding community. The DAO managed to attract nearly 14\% of all ETH tokens in circulation at the time, boasting over 18,000 stakeholders.
Unfortunately, on June 8, 2016, less than three months after its inception, The DAO's smart contract fell victim to a malicious hacker. Over the ensuing weeks, the hacker systematically drained a substantial portion of The DAO's smart contract balance. This security breach dealt a severe blow to The DAO, eroding the trust of its investors and severely denting the credibility of Ethereum and blockchain technology as a whole.
Faced with a formidable decision, the Ethereum core team contemplated potential solutions to thwart the hacker. One option was to execute a fork of the Ethereum blockchain, effectively rewriting its history and creating an alternative reality. By forking Ethereum, the new branch would operate as if the hack had never transpired. If users adopted the new fork and abandoned the old one, the value of the hacker's ETH holdings would significantly decrease. This fork would invalidate the historical blocks containing the hacker's attack transactions. However, this drastic measure ran counter to the fundamental principles underpinning Ethereum. Those who supported the fork were essentially advocating for a world with two parallel Ethereum blockchains. Ultimately, the vote in favor of the fork prevailed with an 85\% majority, leading to the fork's implementation on July 20, 2016, that occurred with block 1,920,000~\cite{fork_block} containing the fix to "The DAO" (i.e., which allowed DAO investors to retrieve their funds). Consequently, two Ethereum chains now exist: Ethereum Classic (which retains the hack in its ledger) and the familiar Ethereum chain we know today (where the ledger's history predates the deployment of the flawed smart contract). Both chains have their native ETH tokens, which possess significantly different market values.

\subsection*{Controversial issues}
These events have sparked two opposing lines of discussion. From a legitimate but unethical perspective, it is essential to delve into the intricacies of this issue. From a purely technical standpoint, the hacker's actions did not breach the parameters established in "The DAO" protocols or the algorithmic rules embedded in the smart contract. This viewpoint gains further weight when considering an open letter signed by the attacker (which a copy can be accessed here~\cite{dao_msg_url}). Nevertheless, the ethical and moral dimensions of this action should not be underestimated. Despite its technical legality, appropriating funds in this manner is regarded as theft, giving rise to a significant ethical and moral dilemma.
On the other hand, in the context of blockchain being an immutable ledger, the outcome leads to a dilemma regarding whether the Immutability Theorem has been compromised due to consensus among network validators. Indeed, the introduction of a hard fork, while addressing the crisis immediately, opens a Pandora's box of philosophical questions that pertain to the very foundations of blockchain technology. It challenges the long-standing principle that code, once implemented, is sacrosanct and akin to law etched in stone, rendering any action permitted by the code inherently legitimate and unalterable once executed. In practical terms, the hard fork operates as a mechanism for temporal regression. Transactions recorded on the public ledger are effectively nullified, creating a reality in which the malicious hack appears to have never happened, as the smart contract was never published on the network. This decision carries profound implications, as it necessitates a compromise on the immutability of the blockchain, a fundamental principle of distributed ledger technology. This compromise is made in the interest of preserving the then-emerging Ethereum movement during a severe existential crisis. The immutable nature of the blockchain, once hailed as a cornerstone principle, is sacrificed in this instance in pursuit of the greater good.

%% file: tex/atm.tex
\section{Understanding Reentrancy}
\label{s:atm}
In the following section we are providing a non-technical explanation of the reentrancy attack using a "bugged" ATM analogy. Imagine you have 10,000 CHF in your bank account, and you walk up to an ATM to withdraw 200 CHF. You receive the 200 CHF, but you notice that your balance hasn't changed. So, you decide to withdraw another 200 CHF, and again, there's no change in your balance. You continue to withdraw increasingly larger amounts until the cash in your hand exceeds your total balance. You keep going, and only when you remove your card does your balance finally reflect what just happened: you now have 0 CHF in your bank balance but 200,000 CHF in your hands. 
All you know is that you now have 200,000 CHF in cash because the ATM kept withdrawing from your original balance without updating it after each withdrawal. Every time you selected "Withdraw 200 CHF," the ATM checked that your balance was sufficient (seeing your original 10,000 CHF balance) and withdrew from it. However, it never updated the balance to 9,800 CHF after each withdrawal. You effectively trapped the ATM in a loop of withdrawing from your initial balance indefinitely, and the money the ATM distributed to you came from the bank's funds, not necessarily your own.
This is precisely what occurred in "The DAO" hack, where a similar vulnerability in The DAO's smart contract code allowed a malicious attacker to drain funds beyond the allocation to which they were entitled. This type of attack is known as a reentrancy attack (or exploit). Just like in the ATM example above, the malicious attacker repeatedly entered a transaction via a recursive call and continuously executed withdrawals without the balance ever being updated. The technical description of this attack is illustrated in the next section.

%% file: tex/dao_code.tex
\section{Technical analysis of the DAO attack}
\label{s:technical_analysis}
\begin{figure}
    \centering
    \includegraphics[width=0.8\linewidth]{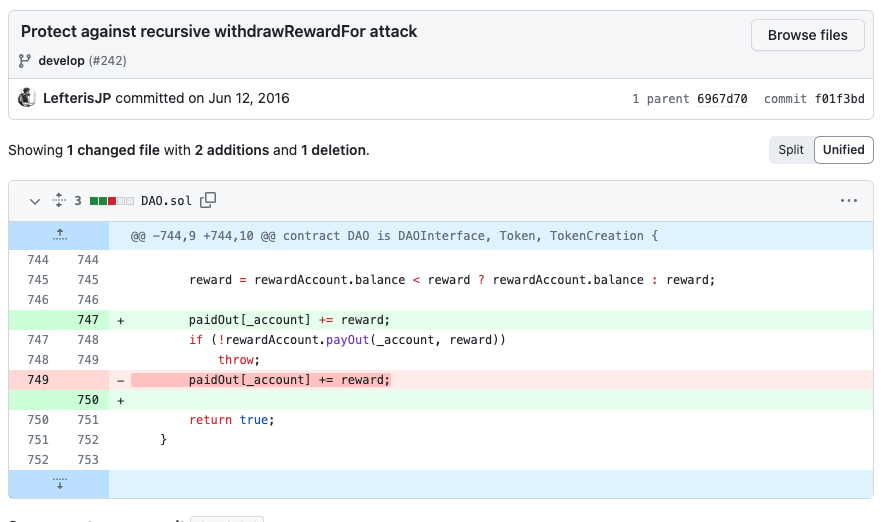}
    \caption{DAO fix github}
    \label{f:dao_fix_github}
\end{figure}

The smart contract "The DAO" is a Solidity-based smart contract (version v0.3.1) consisting of approximately 1200 lines of code, accessible at Ethereum address \texttt{\small 0xbb9bc244d798123fde783fcc1c72d3bb8c189413} (i.e., see~\cite{dao_address}). As previously described, this smart contract was hacked for an amount of 50 million USD on June 17, 2016, by exploiting a flaw in the lines within the \texttt{withdrawRewardFor(..)} function~\cite{dao_sol_url}. This code defect was addressed by Lefteris Karapetsas in the fix titled "Protect against recursive withdrawRewardFor attack"~\cite{dao_sol_fix_url}, by moving the line containing the statement \texttt{paidOut[\_account] += reward;} as depicted in Figure~\ref{f:dao_fix_github}. In essence, what the hacker did was withdraw their previously deposited ETH recursively using the \texttt{splitDAO(..)} function, which invoked the \texttt{withdrawRewardFor(..)} function up to a depth of 29 recursive calls. Consequently, transfers were executed 29 times without incrementing the value of \texttt{paidout[\_account]} with the already paid amount.

\subsection*{A simplified version of The DAO smart contract}
In order to provide a more comprehensive description of how it was possible to exploit "The DAO" smart contract (which was authored by highly experienced individuals) by leveraging the incorrect placement of a single line of code, we simplified the original source code. We rewrote it in Table~\ref{t:dao_source_code} with only the functions necessary to understand its operation and how, by changing the order of just one line, it is possible to alter the transaction outcome. This highlights a fundamental discrepancy between the execution model of Solidity and that on the blockchain. 
\begin{table}[]
    \centering
    \includegraphics[width=0.7\textwidth]{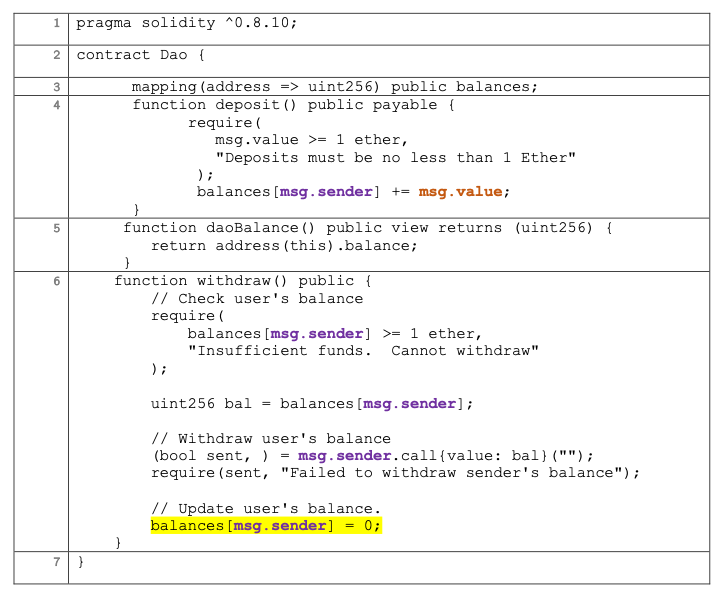}
    \caption{Simplified version of the original "The DAO" smart contract source code.}
    \label{t:dao_source_code}
\end{table}
In the following, we analyze block by block (identified by the numbers in the left column of the table) the various components of this smart contract: 
\textcircled{\small{1}} it identifies the Solidity compiler version used to build the smart contract deployed bytecode.
\textcircled{\small{2}} this line defines the smart contract name (i.e., like a Java class).
\textcircled{\small{3}} this is an internal smart contract state variable that contains the ETH balance value of each mapped address. 
\textcircled{\small{4}} the function \texttt{deposit(..)} is used to deposit some ETH (i.e., defined by \texttt{msg.value}) on the smart contract. This function is used to increment the caller (i.e., identified by \texttt{msg.sender}) balance. By construction requirements, the minimum deposit is 1 ETH. 
\textcircled{\small{5}} the function \texttt{daoBalance(..)} is used to return the available ETH balance stored in the smart contract. 
\textcircled{\small{6}} the function \texttt{withdraw(..)}  is used to withdraw the caller (i.e., identified by \texttt{msg.sender}) balance and it is used to describe in an equivalent manner the functioning of the \texttt{withdrawRewardFor(..)} function~\cite{dao_sol_url} available in the original \texttt{DAO.sol} smart contract. The operations performed by the withdraw function are:
\begin{enumerate}
\item Check if the caller has sufficient funds by checking \texttt{balances[msg.sender]}
\item Withdraw the balance sending the funds to the msg.sender address
\item Update \texttt{balances[msg.sender]} by setting the value to 0
\end{enumerate}
This contract, while appearing straightforward, harbors a significant concern within its \texttt{withdraw(..)} function. The anticipated execution sequence, which aligns naturally with our thought processes when using a sequential language like Solidity, unfolds as follows:
\begin{enumerate}
\item Invocation of the \texttt{withdraw(..)} function.
\item Within the function, a validation step is executed to ascertain the caller's possession of available funds. This validation relies on inspecting the \texttt{balances(address => uint256)} mapping.
\item If the caller possesses available funds, the function proceeds to transfer all those funds back to the caller. Conversely, if no available funds are detected, an error is generated, and the execution terminates.
\item As a final step, the balances mapping is updated to reflect a balance of 0 for the caller's address. 
\end{enumerate}
The pivotal question at this point centers on the locus of the issue. The core concern revolves around the fact that the caller of the \texttt{withdraw(..)} function can either be an external wallet (as in the case of individual users) or another smart contract. 
\begin{figure}
    \centering
    \includegraphics[width=0.5\linewidth]{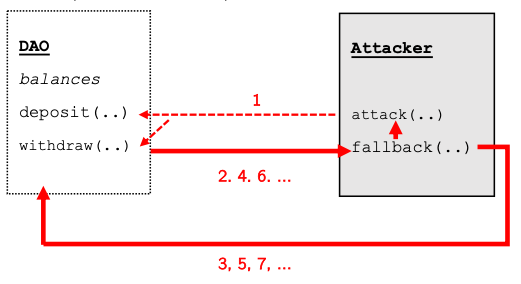}
    \caption{Simplified execution flow of the DAO smart contract attack.}
    \label{f:dao_attack_df}
\end{figure}
In the case the caller is a wallet, the \texttt{withdraw(..)} function operates smoothly without intrinsic issues. In contrast, in the latter scenario where the caller is a smart contract, complications arise due to the non-atomic nature of the function's execution.  The intricacy lies in how this non-atomicity can be exploited by the invoking smart contract.  To elucidate this concept, we employ visual aids of Figure~\ref{f:dao_attack_df}, representing each smart contract as a distinct entity denoted by a box. We further illustrate communication between these smart contracts as communication channels, akin to buffered interconnections. In this diagram  we have two smart contracts communicating with each other: a) the \texttt{DAO} that is the one we saw earlier, and b) \texttt{Attacker} which is the smart contract we use to perform the exploit.
Let us see below how we can build the smart contract \texttt{Attacker} in order to drain all the funds, even those that do not belong to us, from \texttt{DAO} with the few Solidity source code lines illustrated in Table~\ref{t:dao_attack_source_code}. In the following, we analyze block by block (identified by the numbers in the left column of the table) the various components of this smart contract:  
\textcircled{\small{1}} it identifies the Solidity compiler version used to build the smart contract deployed bytecode.
\textcircled{\small{2}} this is the interface of the DAO smart contract, and it is used to define what functions of the DAO smart contract we can call from the Attacker smart contract .
\textcircled{\small{3}} this line defines the smart contract name
\textcircled{\small{4}} this is the handler to the DAO smart contract containing its public address.  
\textcircled{\small{5}} this is the smart contract constructor function, used only during the deployment where the address of the DAO smart contract is provided.  
\textcircled{\small{6}} this is the fallback function, a special Solidity construct that is triggered in specific situations such as when the smart contract receives some ETH.
\textcircled{\small{7}} we implement and use the function \texttt{attack(..)} to launch the attack. We call the deposit function from the DAO smart contract sending it 1 ETH so that: a) it receives the minimum required deposit and b) it records on its balances variable that we have 1 ETH we can withdraw. Finally, we call the \texttt{withdraw(..)} function from the DAO smart contract. This will then send the funds to this smart contract and the fallback function will be triggered and executed. And this is where the problems begin, which we see in detail in the section below.
\begin{table}[]
    \centering
    \includegraphics[width=0.7\textwidth]{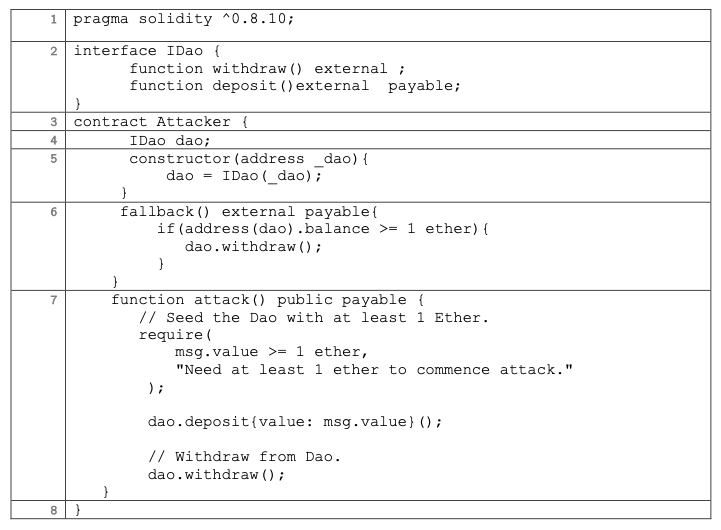}
    \caption{Attacker smart contract Solidity source code.}
    \label{t:dao_attack_source_code}
\end{table}

\subsection*{The reentrancy attack}
\begin{figure}
    \centering
    \includegraphics[width=0.95\linewidth]{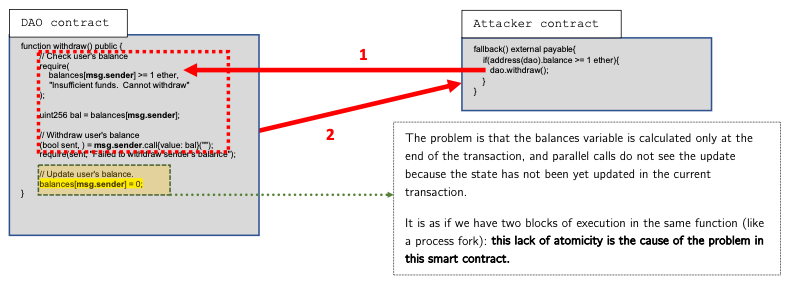}
    \caption{DAO attack explained}
    \label{f:dao_explained}
\end{figure}

Now that we have seen the source code of the Attacker smart contract, let us assume that there are 2 users that we will identify as \texttt{userA} and \texttt{userB} (in reality they are identified by a 42-character hexadecimal address, but this would unnecessarily complicate the discussion). Both \texttt{userA} and \texttt{userB} send 3 ETH in the DAO contract. 
So, we can start the discussion with the DAO contract in the following state:
\begin{itemize}
    \item DAO \texttt{balances[userA]} = 3 ETH
    \item DAO \texttt{balances[userB]} = 3 ETH
    \item The total DAO smart contract balance is 6 ETH (i.e., the total of ETH stored in the smart contract) and this value can be retrieved by the primitive solidity function \texttt{DAO.daoBalance(..)}
\end{itemize}
And now we are ready to launch the attack by calling the \texttt{attack(..)} function from the smart contract Attacker. What happens next is illustrated by the red arrows in Figure~\ref{f:dao_explained}, which is:
\begin{enumerate}
    \item The \texttt{Attacker.attack(..)} is executed and:
    \begin{enumerate}
        \item It calls the \texttt{DAO.deposit(..)} function by sending 1 ETH 
        \item The \texttt{DAO.balances[address(Attacker)]} = 1 ETH is set
        \item It call the \texttt{DAO.withdraw(..)} function
    \end{enumerate}
    \item The \texttt{DAO.withdraw(..)} is called, the value \texttt{DAO.balances[address(Attacker)]} is 1 so:
    \begin{enumerate}
        \item 1 ETH is sent from the DAO contract to the Attacker contract
        \item The new DAO balance is 5 ETH
    \end{enumerate}
    \item The \texttt{Attacker.fallback(..)} function is triggered since 1 ETH is received and this function will call the \texttt{DAO.withdraw(..)} function.
    \item The \texttt{DAO.withdraw(..)} is called, the value \texttt{DAO.balances[address(Attacker)]} is still 1 ETH since it has never been updated: the yellow line we previously highlighted in the Solidity code of DAO smart contract contained in Table~\ref{t:dao_source_code} has not yet been executed. 
    \begin{enumerate}
        \item 1 ETH is sent from the DAO contract to the Attacker contract
        \item The new DAO balance is 4 ETH
    \end{enumerate}
    \item We repeat to point 3 till \texttt{DAO.daoBalance(..)} is 0, i.e., all founds have been drained.
\end{enumerate}
The final status of the DAO contract is the following:
\begin{itemize}
    \item DAO \texttt{balances[userA]} = 3 ETH
    \item DAO \texttt{balances[userB]} = 3 ETH
    \item DAO \texttt{balances[Attacker]} = 0 ETH
    \item The DAO smart contract balance is 0 ETH, since all the founds have been drained:
    \begin{itemize}
        \item address(DAO).balance = 0 ETH
        \item address(Attacker).balance = 6 ETH
    \end{itemize}
\end{itemize}
Now we are going to look in detail at what is going on during the execution of these two smart contracts and why the problem that has led to this exploit in such simple source code is related to the fact that the computation model we are using for Solidity is at the root of the problem. The same conclusions can be extended to the sequential programming languages that are used to develop the smart contracts.
As you can see, the funds from the DAO contract are sent before updating the balances variable, the Attacker \texttt{fallback(..)} function is triggered which will call the DAO \texttt{windraw(..)} function which will have an un-updated view of the balances variable and it will continue to send to Attacker funds that are not its own (DAO owns the funds, but they are intended for userA and userB).

The question that arises is: \textit{"How can we prevent such exploits and attacks?"}
Presently, the prevalent approach employed in smart contract implementation appears to rely on a set of coding best practices. However, this approach poses a substantial challenge as it inherently compromises the security of the code, given the absence of a universally applicable methodology for its analysis, irrespective of the use-case scenario. 
In the next section, we will examine how these techniques are formulated and adopted.
%

%% file: tex/bestpractices.tex
\section{The (fragile and difficult) use of coding best-practices}
\label{s:bestpractices}
\begin{figure}[h]
    \centering
    \includegraphics[width=0.6\linewidth]{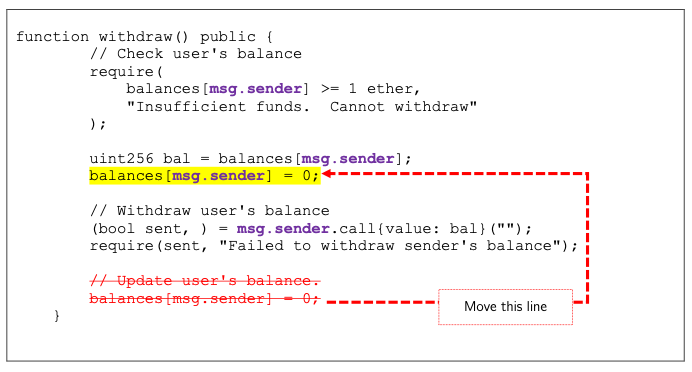}
    \caption{DAO attack reentrancy fix by changing the order of a line.}
    \label{f:dao_fix}
\end{figure}

To date, best practices for smart contracts development are a set of (non-standardized) rules based on the knowledge of experienced developers who suggest some coding rules in order to avoid well-known exploits. There are several collections of best practices for solidity, one of which we believe is the most comprehensive is the one available here~\cite{best_practices}. As you can see, these are \textit{alchemical} rules that look almost hobbyist (even if drafted by a company). For our particular example, the best-practice we need to use in order to prevent the reentrancy attack in our example is the following one: 
\\~\\
\texttt{“If no internal state updates happen after an ether transfer or an external function call inside a method, the function is safe from the reentrancy vulnerability”. }
\\~\\
This rule requires to change the order of operations in the DAO \texttt{withdraw(..)} function so that the caller’s balance is reset to 0 before some ETH are sent to the Attacker smart contract. The new code would look like the one illustrated in Figure~\ref{f:dao_fix}. This is feasible for this simple example but could be not for more complex ones. It's worth noting that this corresponds precisely to the fix that was implemented in the original source code in 2016, as previously illustrated in Figure~\ref{f:dao_fix_github}. A second solution, could be use a kind-of \textit{mutex variable}~\cite{mutex} as the one illustrated in Figure~\ref{f:sc_gen_source} and discussed in the following section.

\subsection*{General considerations about using best practices}
Security-by-design is a fundamental requirement of any technology today. However, with this simple example, we have seen how even the development of a simple smart contract of a few lines is prone to potentially catastrophic bugs.
We have seen how it is extremely difficult to effectively write even simple smart contracts that are secure and reliable since the development methodology is still based on best-practices that, as you can realize, are difficult to apply when it comes to more complicated protocols that require writing high numbers of lines of code or when not applicable in specific situations.
If we go back to the origin of the problem, the main difficulty in securely writing a smart contract can be related to the divergence between the programming language execution model and the real execution on the blockchain. In fact, this fact can be identified as the main cause of problems in both development and analysis. 
\begin{enumerate}
    \item The difficulty in using inappropriate language is evident if you have a look to the log on github of the various attempts by the DAO development team to fix the bug~\cite{dao_sol_fix_url_2}. It must be noted that at that time no best practices where available.
    \item The difficulty in analysing smart contracts developed with an inappropriate execution model is evident when we try to analyse the above example with the current tools: the code of the second solution (i.e., the one with the mutex) generates false positives making the analysis unrealisable and the security of the implementation still dependent to the developer experience.
\end{enumerate}
Therefore, it is absolutely necessary to have a development methodology capable to provide the security by design, which is a fundamental requirement for any kind of technology we have today.
In the following section we will see how the security by design requirement can be granted by using a dataflow programming model and how, in our opinion, pursuing research in this direction is the right direction.

%% file: tex/dataflow.tex
\section{The use of a Dataflow model}
\label{s:dataflow}

\begin{figure}
    \centering
    \includegraphics[width=0.5\linewidth]{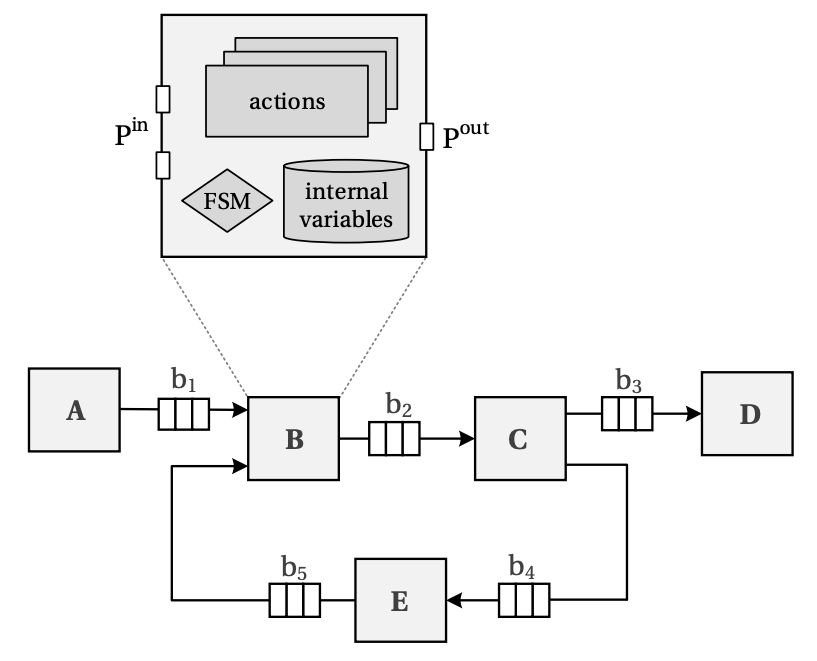}
    \caption{An example of RVC-CAL dataflow network composed by 5 actors (A, B, C, D, E) and 5 buffers (b1, b2, b3, b4, b5). Each actor is composed by a set of input and output ports, a set of internal state variables, a set of actions (i.e., atomic functions) and a finite state machine (FSM).}
    \label{f:cal}
\end{figure}

In recent decades, the rise of massively parallel architectures, coupled with the challenges of programming these architectures, has made the dataflow paradigm an attractive alternative to the imperative paradigm~\cite{CastrillonBook,casale2016programming}. The primary advantages of the dataflow paradigm are linked to its capacity to express concurrency without intricate synchronization mechanisms. This capability arises from the program's internal representation as a network of processing blocks that exclusively communicate through communication channels. In fact, these blocks operate independently and do not produce any side effects~\cite{lee_2001}. Consequently, this eliminates potential concurrency issues that may emerge when programmers are tasked with manually managing synchronization among parallel computations. Furthermore, this paradigm explicitly exposes all the inherent parallelism within a program.
Over the past decade, a multitude of programming languages has emerged to model the semantics of dataflow programs~\cite{scb_thesis,Johnston_2004_advances}. Imperative programming languages have been extended to incorporate parallel directives (e.g., Java, Python, C/C++), while native dataflow languages (e.g., Esterel, Ptolemy) have been newly specified. Within this diverse landscape of language extensions, RVC-CAL~\cite{iso230014} distinguishes itself as the sole formally standardized programming language aligned with the dataflow model. It is capable of modeling complex and dynamic dataflow networks where the token production and consumption rates cannot be known at compile time. As depicted in Figure~\ref{f:cal}, an RVC-CAL actor is defined as a collection of atomic methods (i.e., functions), referred to as actions, accompanied by encapsulated state variables. These variables are inaccessible for access or modification by neighboring actors within the same network. During an actor's execution, only a single action is selected at any given time, with the concurrent execution of multiple actions being precluded. The selection of the action to execute is contingent upon the input token values and/or the actor's internal variables.
One of the intriguing properties associated with the use of these high-level dataflow programming models is the ability to generate optimized and secure-by-design low-level code from this architecture-independent representation. Examples of synthesis tools specifically developed for the RVC-CAL programming language include the Open RVC-CAL Compiler (Orcc)~\cite{Yviquel_Orcc}, Exelixi~\cite{eb_thesis}, and Tÿcho~\cite{cedersjo2019tycho}.

\subsection*{Dataflow-based smart contracts}
\begin{figure}
    \centering
    \includegraphics[width=0.95\linewidth]{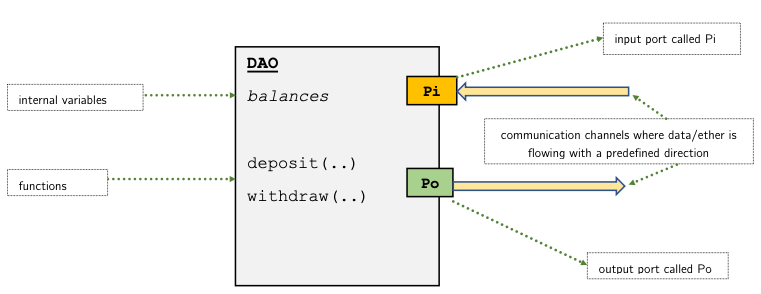}
    \caption{Smart contract dafaflow model}
    \label{f:sc_dataflow}
\end{figure}

The interconnected block representation we employed in Figure~\ref{f:dao_attack_df} to elucidate the interaction between two smart contracts, namely the DAO and the Attacker, essentially forms the foundation of any dataflow-based model, as seen earlier. As illustrated in Figure~\ref{f:sc_dataflow}, the core characteristics of such a dataflow model are as follows:
A) Each box represents an actor, which corresponds to a smart contract, as previously demonstrated in Figure~\ref{f:dao_attack_df}.
B) The exchange of information between two smart contracts can only occur through dedicated communication channels known as buffers. These buffers ensure that the order and consistency of data are preserved and guaranteed by the execution model itself.
C) Each function execution is inherently atomic. This means that, in advance, we are assured that the execution model prevents unpredictable effects resulting from changing the order of source code lines. 
A function execution adheres to the following sequence of operations:
\begin{itemize}
    \item Input data is consumed from the input buffer(s).
    \item Subsequently, the execution occurs, during which state variables may be updated.
    \item Only at this point is output data placed in the output buffer(s).
\end{itemize}
The complexity of the individual action determines the category to which each smart contract (actor) belongs. These categories include:
\begin{itemize}
    \item Static: At each function execution, it consumes from its input buffers and produces on its output buffers an always equal amount of data.
    \item Cyclo-Static: The number of data consumed and produced varies from run to run but follows a repetitive and cyclic pattern.
    \item Dynamic: The number of produced and consumed data is not known in advance.
\end{itemize}
These fundamental rules underlie a dataflow programming model, which can be extended to enhance the expressiveness and capability of representing various smart contract use cases.
Consequently, creating a smart contract using a Domain Specific Language (DSL) similar to Solidity and RVC-CAL can resemble the example presented in Table~\ref{t:sc_df_code}. It is important to note that in this example, we employ a guard condition as a prerequisite for executing the action (function). If the prerequisite is not met, the action cannot be executed.
\begin{table}[]
    \centering
    \includegraphics[width=0.7\textwidth]{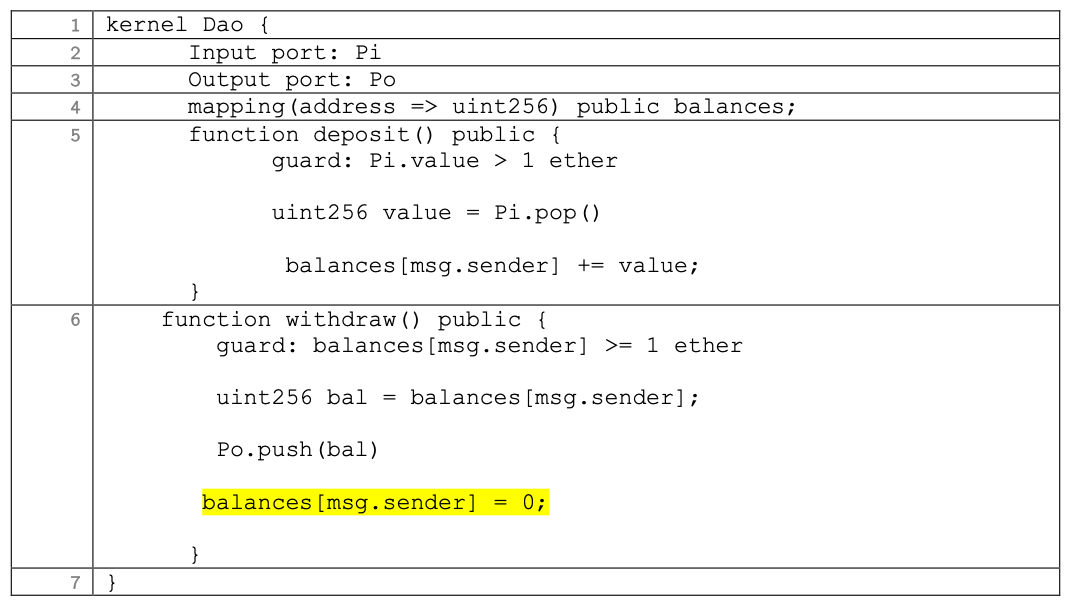}
    \caption{Smart contract dataflow code using a DSL similar to Solidity and RVC-CAL}
    \label{t:sc_df_code}
\end{table}
In the following sections, we will analyze each block, identified by the numbers in the left column of the table, to understand the various components of this smart contract:
\textcircled{\small{1}}  Smart contract (actor) name. 
\textcircled{\small{2}}  Input port definition, the point where data can be read/received inside the actor.
\textcircled{\small{3}} Output port definition, the point where data can be written/sent outside the actor.
\textcircled{\small{4}} ETH balance of each mapped address.
\textcircled{\small{5}}A requirement for at least 1 ether available in the input port, involving popping (consuming) one value from the input port and updating the state variable.
\textcircled{\small{6}} It's worth noting that the update of the balances variable occurs after the value is pushed onto the output port. However, it's crucial to understand that this message will be sent only once all operations are executed. This design choice fundamentally guarantees the absence of a reentrancy condition.

\begin{figure}
    \centering
    \includegraphics[width=0.95\linewidth]{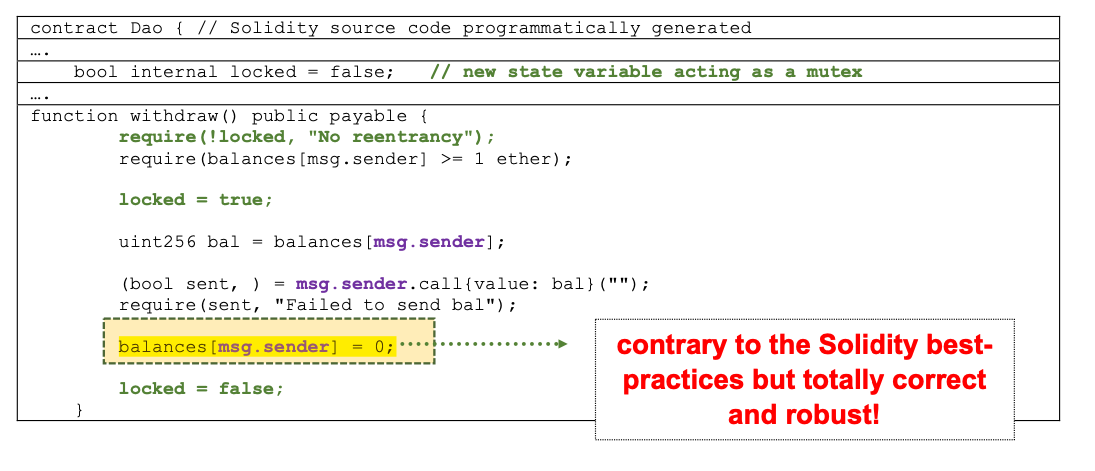}
    \caption{Smart contract dafaflow generated source code}
    \label{f:sc_gen_source}
\end{figure}

In essence, this model enables the generation of Solidity source code that is correct by construction. As previously mentioned, the dataflow model of computation hinges on the concept of atomicity in function execution. This concept can be translated into automated source code generation from the dataflow model to Solidity code, introducing, for example, a mutex variable, as seen in the generated source code illustrated in Figure~\ref{f:sc_gen_source}. It's important to note that this source code, while deviating from Solidity best practices, is entirely correct, and it fundamentally enhances robustness by being invariant to the order of execution of its operations. This assures a security by design property, a critical requirement for smart contract development, which is currently lacking.

%% file: tex/conclusions.tex
\section{Conclusions}
\label{s:conclusions}
In summary, this study represents an initial exploration into the integration of dataflow programming models and Domain-Specific Languages (DSLs) within the domain of smart contract development, with a primary focus on enforcing security through the principle of security-by-construction.
Our investigation has unveiled the fundamental attributes of dataflow-based models, demonstrating their innate capacity to articulate concurrency in a manner that obviates the need for intricate synchronization mechanisms. This approach effectively mitigates potential concurrency-related vulnerabilities, a significant concern within decentralized applications. Moreover, dataflow models provide a transparent framework for exposing inherent program parallelism, imparting an additional layer of security to the smart contract development process.
As elucidated in our analysis, the adoption of DSLs analogous to Solidity and RVC-CAL facilitates the automated generation of low-level code that adheres to security-by-design principles, starting from high-level, architecture-agnostic representations. This approach reframes smart contract development, shifting the emphasis away from manual coding practices, which heavily rely on developer expertise, towards a systematic, inherently secure methodology.
While our research marks an initial exploration of these concepts, it simultaneously beckons forth a promising trajectory for future research. Within the ever-evolving landscape of blockchain technology, where security stands as a paramount concern, this nascent study lays the foundational groundwork for a more resilient and secure future in smart contract programming. 